\documentclass{moriond}
\usepackage{bm}  

\def\be{\begin{equation}}
\def\ee{\end{equation}}
\def\bea{\begin{eqnarray}}
\def\eea{\end{eqnarray}}

\newcommand{\Photo}{\vspace{0.2cm}
\includegraphics[height=35mm]{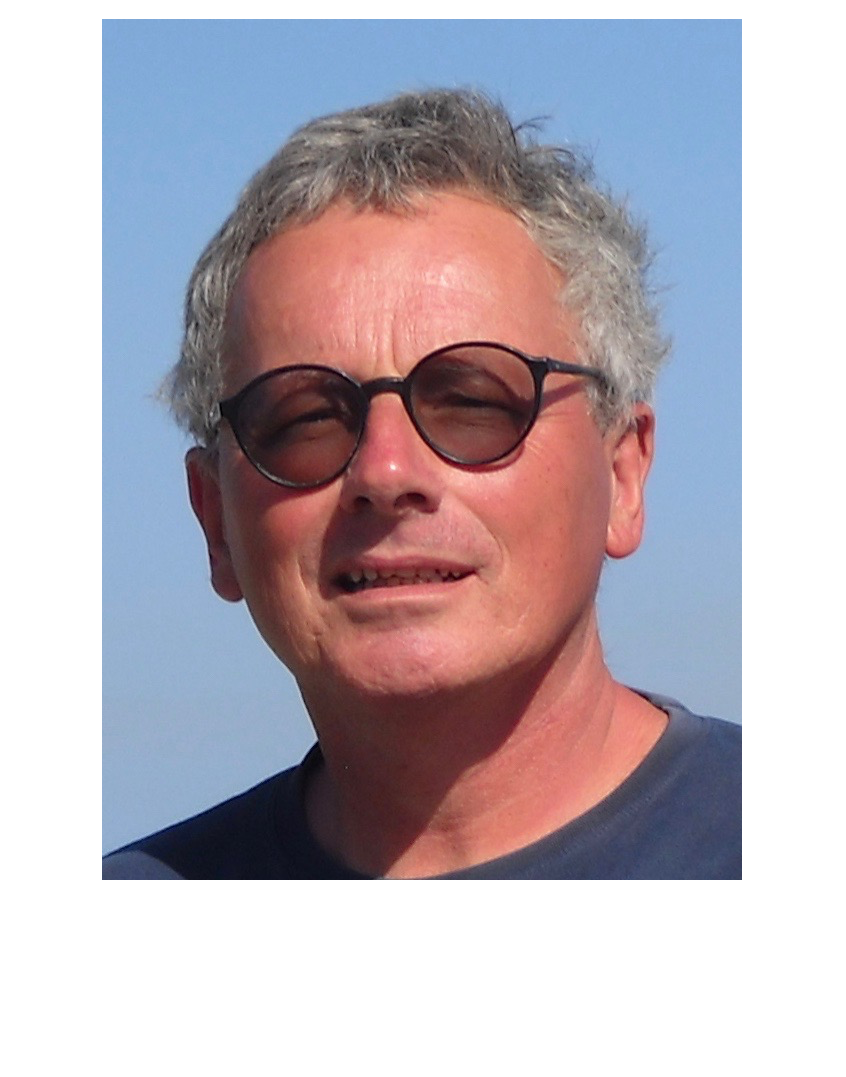}}

\begin{document}
\vspace*{4cm}
\title{EXOTIC INTERPRETATION OF ${\bf \Omega_c}$ EXCITED STATES}

\author{ M. PRASZA{\L}OWICZ }

\address{M. Smoluchowski Institute of Physics, Jagiellonian University, \\ S. {\L}ojasiewicza 11,
30-348 Krak{\'o}w, Poland}

\maketitle

\vspace*{-1.1cm}
\abstracts{
We use the chiral quark-soliton model to interpret five excited $\Omega_c$ states
recently reported by the LHCb collaboration and confirmed by Belle. We briefly recapitulate
the model and its application to light baryons. We then show how the model can be extended
to the case of baryons with one heavy quark. We test the model against ground state heavy
baryons and then examine possible excitations. We argue that it is
not possible to accommodate all five $\Omega_c$'s within five parity minus excitations predicetd
by the model and
propose to interpret two narrowest states split by 70 MeV as pentaquarks belonging to
the SU(3) representation $\overline{15}$. }

\section{Chiral Quark Soliton Model ($\chi$QSM)}

$\chi$QSM \cite{Diakonov:1987ty} (for  review see Ref.~\nocite{Christov:1995vm}[2] and references therein)
 is based on an old argument by Witten, which says that in the limit of a large number of colors ($N_c \rightarrow \infty$), 
 $N_c$ relativistic valence quarks generate chiral mean fields represented by a distortion of 
 a Dirac sea that in turn interacts with the valence quarks themselves.
The soliton configuration corresponds to the solution of the Dirac equation for the constituent quarks (with
gluons integrated out) in the mean-field approximation, where pseudoscalar mean fields respect so called {\em hedgehog}
symmetry, since it is impossible to construct a pseudoscalar field that changes sign under
inversion of coordinates, which would be compatible with
the SU(3)$_{\rm flav}\times$SO(3) space symmetry. 
This means that 
neither spin ($\bm{S}$) nor isospin ($\bm{T}$)  are  {\em good} quantum numbers. Instead a {\em grand spin}  
$\bm{K}=\bm{S}+\bm{T}$ is a {\em good} quantum number.

\begin{figure}[h]
\centering
\includegraphics[width=9.0cm]{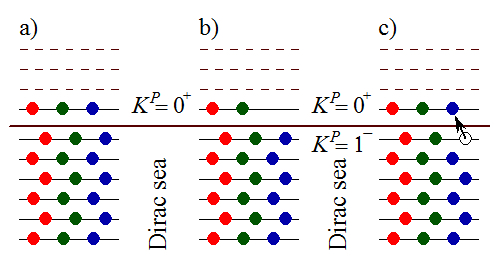} \vspace{-0.2cm}\caption{Schematic
pattern of light  quark levels in a self-consistent soliton
configuration. In the left panel all sea levels are filled and $N_{c}$ (=3 in
the Figure) valence quarks occupy the $K^{P}=0^{+}$ lowest
positive energy level. Unoccupied positive energy levels are dpicted
by dashed lines. In the middle panel one valence quark has been
stripped off, and the soliton has to be 
supplemented by a heavy quark not shown in the Figure. In the right panel a
possible excitation of a sea level quark, conjectured to be $K^{P}=1^{-}$, to
the valence level is shown, and again the soliton has to couple to a heavy
quark. Strange quark levels that exhibit different filling pattern are
not shown.}%
\label{fig:levels}%
\end{figure}

The ground state configuration corresponds to the fully occupied $K^P=0^+$ valence level, as shown in Fig.~\ref{fig:levels}.a.
Therefore the soliton does not carry definite quantum numbers except for the baryon number resulting from the valence quarks.
Spin and isospin appear when the rotations in space and flavor are quantized 
and the resulting {\em collective} hamiltonian is analogous to the one
of a symmetric top. There are two conditions that the {\em collective} wave funcions
have to satisfy:
\begin{itemize}
\item allowed SU(3) representations must contain states with hypercharge
$Y^{\prime}=N_{c}/3$,
\item the isospin $\bm{T}^{\prime}$ of the states with $Y^{\prime}%
=N_{c}/3$ couples with the soliton spin $\bm{J}$ to a singlet:
$\bm{T}^{\prime}+\bm{J}=0$.
\end{itemize}
As a result, the lowest praity (+) baryons belong to the SU(3)$_{\rm flavor}$ octet of spin 1/2 and
decuplet of spin 3/2. The first {\rm exotic} representation is  $\overline{\mathbf{10}}$ of spin 1/2 
with the lightest state corresponding to
the putative $\Theta^{+}(1540)$  (see {\em e.g.} Moriond proceedings~\cite{Moriond2005} 2005).
The model has been successfully tested in the light baryon sector.

\section{$\chi$QSM and heavy baryons}

\begin{figure}[h]
\centering
\includegraphics[height=5cm]{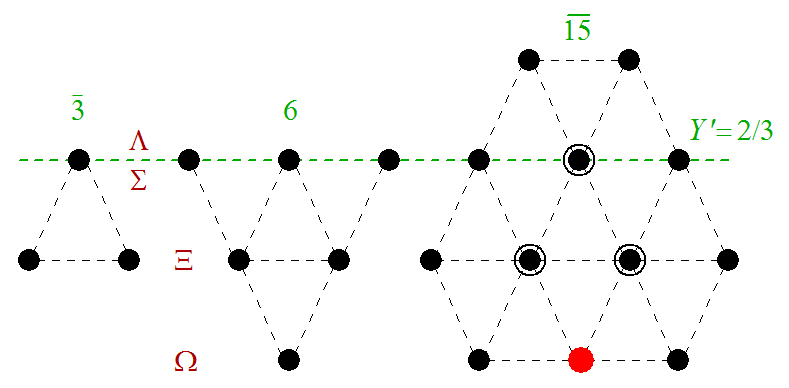}
\caption{Rotational
band of a soliton with one valence quark stripped off. Soliton spin
corresponds to the isospin $T^{\prime}$ of states on the quantization line
$Y^{\prime}=2/3$. We show three lowest allowed representations: antitriplet of
spin 0, sextet of spin 1 and the lowest exotic representation $\overline
{\mathbf{15}}$ of spin 1 or 0.  Heavy quark has to be added.}
\label{fig:irreps}%
\end{figure}

Recently wa have proposed~\cite{Yang:2016qdz}, following Ref.~[6]\nocite{Diakonov:2010zz} to generalize the above approach to heavy baryons,
by stripping off one valence quark from the $K^P=0^+$ level, as shown in Fig.~\ref{fig:levels}.b,
and replacing it by a heavy quark to neutralize the color. 
In the large $N_c$ limit both systems: light and heavy baryons are described essentially by the same mean field, and the
only difference is now in the quantization condition:
\begin{itemize}
\item allowed SU(3) representations must contain states with hypercharge
$Y^{\prime}=(N_{c}-1)/3$.
\end{itemize}
The lowest allowed SU(3) representations are in this case
 $\overline{\mathbf{3}}$ of spin 0  and 
to ${\mathbf{6}}$ of spin 1 shown in Fig.~\ref{fig:irreps}.

\begin{figure}[h]
\centering
\includegraphics[height=7cm]{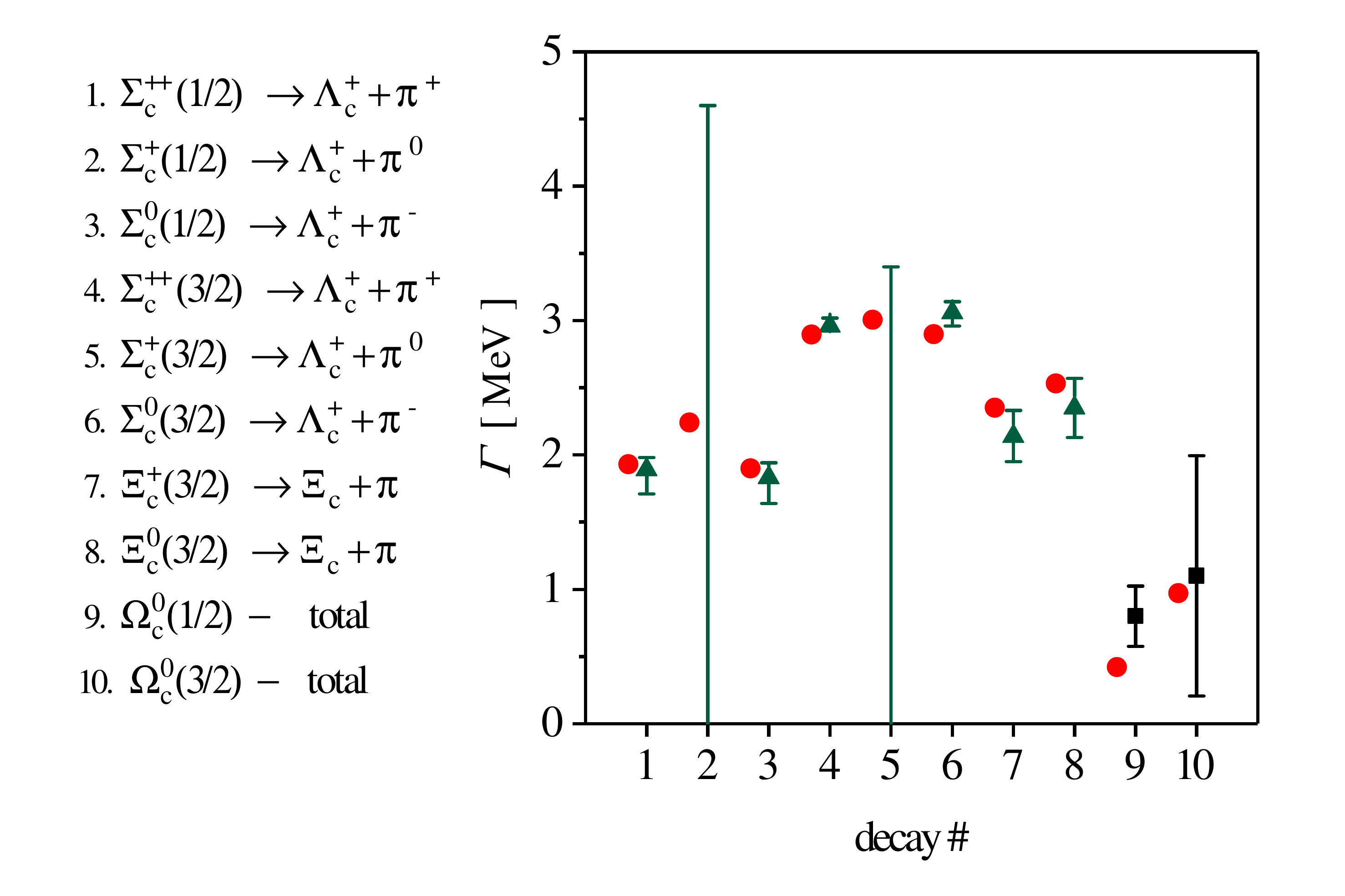}
\caption{Decay
widths of the charm baryons.  Red full circles correspond to our
theoretical predictions. Dark green triangles correspond to the experimental
data$^7$ 
Data for decays 4 -- 6 of 
$\Sigma_{c}(\mathbf{6}_{1},3/2)$ have been divided by a factor
of 5 to fit within the plot area. Widths of two LHCb$^8$ 
$\Omega_{c}$ states that we interpret as pentaquarks are plotted as black full
squares with  theoretical values shown as red full circles. }%
\label{fig:charmwidths}%
\end{figure}

An important feature of this approach is that
both  ${\mathbf{6}}-\overline{\mathbf{3}}$ splitting and the splittings inside
these multiplets due to the strange quark mass are {\em predicted} using as an input
the light sector spectrum and are in good agreement with experiment~\cite{Yang:2016qdz}. The new 
ingredient is a hyperfine splitting due to the spin-spin interaction of a soliton
and a heavy quark, which can be parametrized phenomenologically. Moeover,
the decay widths can be calculated within the same approach, and the results for
the charm baryons are shown  in Fig.~\ref{fig:charmwidths}.

\section{Excitations of heavy baryons}

Two possible kinds of excitations are present in the $\chi$QSM. Firstly, higher
SU(3) representations, similar to the antidecuplet in the light sector, appear
in the rotational band of the 
soliton of Fig.~\ref{fig:levels}.b.
The lowest possible exotic SU(3) representation is 
$\overline{\mathbf{15}}$ of positive parity  and spin 1
 ($\overline{\mathbf{15}}$ of spin 0 is heavier) 
depicted in Fig.~\ref{fig:irreps}. Second possibility corresponds to the
excitation of the sea quark from the $K^P=1^{-}$ sea level to the valence level~\cite{Diakonov:2010zz}
shown in Fig.~\ref{fig:levels}.b (or alternatively valence quark excitation to the first excited level~\footnote{We thank Victor
Petrov for pointing out this possibility.}
of $K^P=1^{-}$). In this case the parity is negative but the rotational band is the same 
 (see Fig.~\ref{fig:irreps}) with, however, different quantization condition:
\begin{itemize}
\item the isospin $\bm{T}^{\prime}$ of the states with $Y^{\prime}%
=(N_{c}-1)/3$ couples with the soliton spin $\bm{J}$ as follows:
$\bm{T}^{\prime}+\bm{J}=\bm{K}$, where $\bm{K}$ is the grand spin
of the excited level.
\end{itemize}

\begin{figure}[h]
\centering
\includegraphics[width=9.0cm]{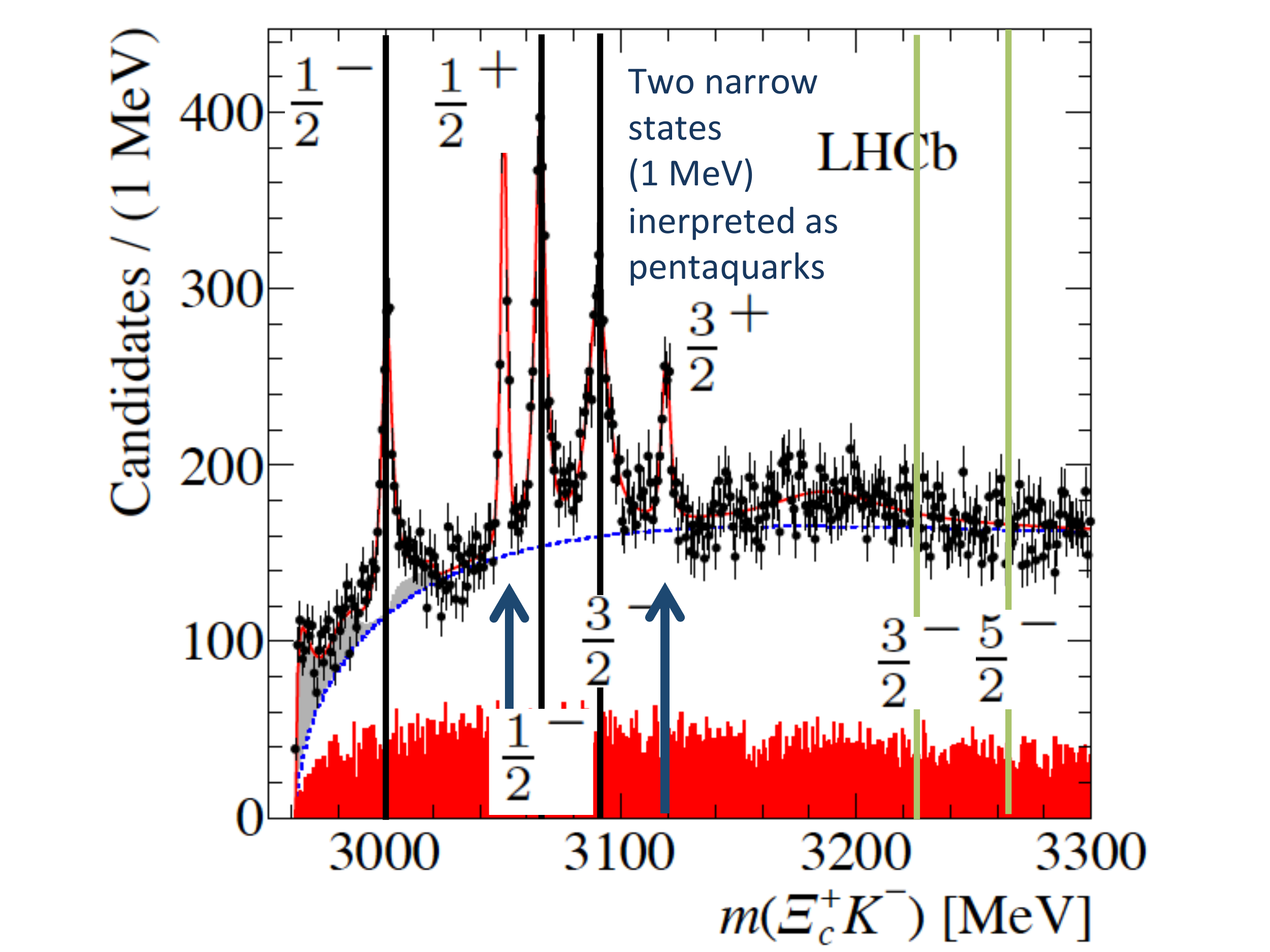} \vspace{-0.2cm}
\caption{Spectrum of the $\Omega^0_c$ states (from Ref.[8]) with theoretical predictions of the present model}%
\label{fig:Omegas}%
\end{figure}

We have shown that the model describes well the only fully known spectrum of negative parity antitriplets
of spin 1/2 and 3/2~\cite{Yang:2016qdz}. There has been no experimental evidence for the sextet until recent
report of five $\Omega^0_c$ states reported by the LHCb~\cite{Aaij:2017nav} and confirmed by BELLE~\cite{Yelton:2017qxg}.
In the sextet case the above mentioned condition predicts that the soliton spin can be quantized
as $J=0,1$ and 2. By adding one heavy quark we end up with five possible total spin $S$ excitations: for $J=0$ $S=1/2$, for $J=1 $ $S=1/2$ and 3/2, 
and for $J=2$  $S=3/2$ and 5/2. Although the number of states coincides with the experimental results~\cite{Aaij:2017nav,Yelton:2017qxg},  it is
not possible to accommodate all five $\Omega^0_c$ states within the constraints imposed by the $\chi$QSM~\cite{Yang:2016qdz}. We have
therefore {\em forced} model constraints (note that in the $\mathbf{6}$ case we cannot predict the mass splittings, since there is a new 
parameter in the splitting hamiltonian that corresponds to the transition of Fig.~\ref{fig:levels}.c, which is not known from the light sector),
which allows to accommodate only three out of five LHCb states (see black vertical lines in Fig.~\ref{fig:Omegas}). 
Two heaviest $\chi$QSM states (green lines in  Fig.~\ref{fig:Omegas}) lie already above the decay threshold to heavy mesons,
and it is quite possible that they have very small branching ratio to the  $\Xi^+_c+K^-$ final state analyzed by the LHCb. Two remaining states
indicated by dark-blue arrows in Fig.~\ref{fig:Omegas}, which are 
hyper fine  split by 70~MeV  (as the ground state sextets that belong to the same rotational band), 
can be therefore interpreted as the members of exotic 
$\overline{\mathbf{15}}$ of positive parity shown as a red dot in Fig.~\ref{fig:irreps}. This interpretation is reinforced
by the decay widths, which can be computed in the model. These widths are of the order of 1~MeV and agree with the
LHCb measurement (see Fig.~\ref{fig:charmwidths}). Such small widths are in fact expected in the present approach,
since the leading $N_c$ terms of the relevant couplings cancel in the non-relativistic limit.

The simplest way to falsify or to confirm our identification is to search 
 for the {\it isospin} partners of $\Omega^0_c$ from the $\overline{\mathbf
   15}$. They can be searched in the mass distribution of
 $\Xi_c^0+K^-$ or $\Xi_c^+ + \bar K^0$: the $\Omega^0_c$'s from the
 sextet do not decay into these channels. Our model applies also to the bottom sector,
 and -- where the data is available -- it describes very well both masses and decay widths.

\section*{Acknowledgments}
It is a pleasure to thank my collaborators H.-C.~Kim, M.V.~Polyakov and G.S.~Yang 
and the organizers of the Rencontres de Moriond for giving me an opportunity to present
our work.

\end{document}